\documentclass[aps,prl,reprint,superscriptaddress]{revtex4-2}

\usepackage[a4paper, total={174mm,259mm},left=18mm,top=18mm]{geometry}
\usepackage{amsmath,amssymb,amsfonts}
\usepackage{algorithmic}
\usepackage{graphicx}
\usepackage{wrapfig}
\usepackage{caption}
\usepackage{subcaption}
\usepackage{float}
\usepackage{hyperref}
\hypersetup{colorlinks=true,linkcolor=teal,urlcolor=blue,filecolor=red,citecolor=olive}
\usepackage{mathrsfs}
\usepackage{tcolorbox}

\begin{document}
\title{Low-temperature benchmarking of qubit control wires by primary electron thermometry}
%\title{Benchmarking with Primary Electron Thermometry}
%\author{Elias Roos Hansen, Joost van der Heijden, Ferdinand Kuemmeth}

\author{Elias Roos Hansen}
\affiliation{Center for Quantum Devices, Niels Bohr Institute, University of Copenhagen, 2100 Copenhagen, Denmark}

\author{Ferdinand Kuemmeth}
\affiliation{Center for Quantum Devices, Niels Bohr Institute, University of Copenhagen, 2100 Copenhagen, Denmark}
\affiliation{QDevil, Quantum Machines, 2750 Ballerup, Denmark}

\author{Joost van der Heijden}
\affiliation{QDevil, Quantum Machines, 2750 Ballerup, Denmark}

\begin{abstract}
    Low-frequency qubit control wires require non-trivial thermal anchoring and low-pass filtering.
    The resulting electron temperature serves as a quality benchmark for these signal lines.
    In this technical note, we make use of a primary electron thermometry technique, using a Coulomb blockade thermometer, to establish the electron temperature in the millikelvin regime.
    The experimental four-probe measurement setup, the data analysis, and the measurement limitations are discussed in detail.
    We verify the results by also using another electron thermometry technique, based on a superconductor-insulator-normal metal junction.
    Our comparison of signal lines with QDevil’s QFilter to unfiltered signal lines demonstrates that the filter significantly reduces both the rms noise and electron temperature, which is measured to be 22 $\pm$ 1 mK.
\end{abstract}

\maketitle

\begin{figure*}[!ht]
    \centering
    \includegraphics[width=1.00\linewidth]{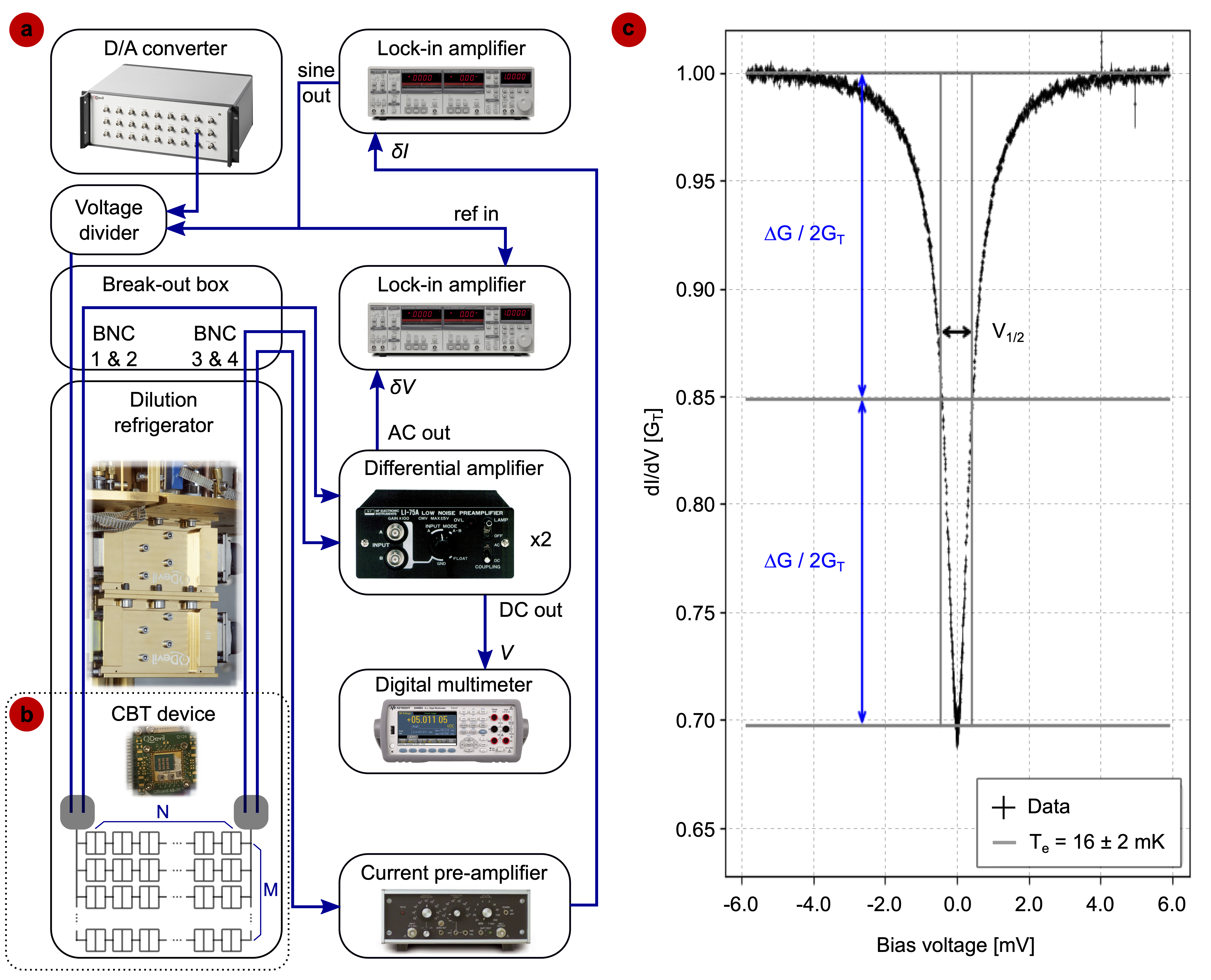}
    \captionof{figure}{(a) \textit{Experimental four-probe setup for differential conductance measurement of a CBT device inside a dilution refrigerator, including: QDevil QDAC digital-analog converter; Ithaco 1211 current pre-amplifier; Keysight 34465A digital multimeter; Li-75A low-noise pre-amplifiers; Stanford Research Systems SR800-series lock-in amplifiers; QDevil QBox breakout box; and a resistive voltage divider.} (b) \textit{Schematic of a CBT device, consisting of $M$ parallel arrays of $N$ tunnel junctions in series.} (c) \textit{Measured differential conductance around zero bias of a CBT device with $N=100$ and $M=10$. The characteristic dip as a function of bias voltage depends only on well-known physical constants and the electron temperature. $\Delta G/G_T$ and $V_{1/2}$ are extracted to find an electron temperature T$_e$ = 16 $\pm$ 2 mK, using equation \ref{EqCBT2}. The uncertainty is found by repetition measurements. Note that this value underestimates T$_e$ due to a bias-heating effect (see Appendix \ref{subsec:biasheating}), and a more accurate estimation is found to be 22 $\pm$ 1 mK.}}
    \label{fig:1}
\end{figure*}

Cryogenic experiments are essential for quantum science, as a low electron, phonon or photon temperature is crucial for the emergence of many quantum properties and for the performance of associated quantum technologies. With the use of dilution refrigerators, \emph{phonon} temperatures of the order of 10 millikelvin are routinely generated in laboratories. In practice, however, other relevant temperatures in an experiment (such as effective \emph{electron} temperature) can be difficult to determine for a number of reasons. Noise levels may exceed the thermal energy, low-temperature thermal conductivities are generally low, resulting in inhomogeneous and time-dependent temperatures due to variations in local dissipation or local heat loads, and different physical subsystems (conduction electrons, nuclear spins, phonons, electro-magnetic modes in the microwave regime, magnetic excitations, etc.) may thermally decouple, resulting in long equilibration times. (The intrinsic heat leak arising from the ortho-para conversion of hydrogen molecules in the cryostat's copper structures, for instance, may take weeks~\cite{Pobell2007}.) 
In particular, the coupling between conduction electrons and phonons vanishes at low temperatures, resulting in two subsystems that are generally out of mutual equilibrium and must be characterized individually by their own effective temperatures \cite{Wellstood}.

Due to low electron-phonon coupling, minimizing electron temperatures in a dilution refrigerator requires non-trivial thermal anchoring of the signal wires as well as careful low-pass filtering to reduce the thermal effects of noise over a large bandwidth (tens of GHz). Furthermore, the resulting temperatures of phonons and electrons in a sample holder must be established separately and generally require different techniques. In this application note, we address the challenge of determining the effective electron temperature of low-frequency control wires inside a cryogenfree dilution refrigerator by using metallic Coulomb blockade thermometers (CBTs) and by comparing this technique to the use of superconductor-insulator-normal metal (SIN) junctions as thermometers. For a particular cryostat configuration, with the samples loaded via a bottom-loading mechanism onto the cold-finger inside a vector magnet, we determine an electron temperature of 22 ± 1 mK with the CBT method and 27 ± 1 mK with the SIN method, while the temperature of the sample holder itself, measured resistively via a calibrated ruthenium-oxide (RuO$_2$) thermometer, was indicated at 14 mK. These effective temperatures can be understood as an upper bound of the electron temperature, and are similar to results obtained from semiconducting-quantum-dot thermometry in a previous study~\cite{QDevil2021}. Based on the CBT and SIN data, we analyse the effect of QDevil's cryogenic filters compared to unfiltered reference wires. We hope that this application note is useful for those wishing to test cryogenic setups for the use of qubit control or other applications where low electron temperature and low noise levels in low-frequency control wires is important. 

\section{Background}\label{sec:background}
A CBT is a primary thermometer designed to be insensitive to environmental influences, including variations in external magnetic fields \cite{Pekola2004}, offset-charging effects \cite{Yur}, and fabrication errors \cite{pekola2021influence}. This method was first introduced in the 1990's~\cite{pekola1994} and has been further refined over the last years \cite{intermediate, Pekola2004, pekola2021influence, JoonasRef3}. Here, we provide a practical introduction to performing and analysing CBT measurements, illustrated by measurements obtained in our lab.

The CBT, depicted in the drawing of Fig.~\ref{fig:1}b, consists of arrays of metallic islands separated by tunnel junctions. The thermometer operates in the regime where the charging energy of the islands, $E_c\propto 1/C$, determined by each island's total capacitance $C$, is of the order of or smaller than the thermal energy, $E_c\lesssim k_BT_e$, where $T_e$ is the electron temperature and $k_B$ is Boltzmann's constant~\cite{Yur}. (This is the opposite limit of the single electron transistor \cite{intermediate}, which can be used as a secondary thermometer as outlined in an earlier QDevil application note~\cite{QDevil2021}.) Under such circumstances, the differential conductance is expected to show a 'universal dip' around zero voltage bias \cite{Czech, Pekola2004}, as shown in the measurement of Fig.~\ref{fig:1}c.

In the appropriate operating regime, the relative shape of the conductance dip depends exclusively on the electron temperature and a number of physical constants. In particular, in the lowest-order estimate~\cite{intermediate} the full-width-half-maximum (FWHM) is given by:
\begin{equation}\label{EqCBT1}
    V_{1/2}\approx5.439k_BNT_e,
\end{equation}
where $N$ is the number of junctions. This means that the temperature reading requires no calibration~\cite{Pekola2004,intermediate}.

\section{Method}\label{sec:method}
Figure~\ref{fig:1}a illustrates the main components of an experiment dedicated to determine the electron temperature using a CBT device. Our device has $N=100$ aluminum islands in series and $M=10$ junction arrays connected in parallel. More details on the specific device design can be found in Ref.~\cite{OurSensor2011}.

\subsection*{Setup}\label{subsec:setup}
The CBT device is wire-bonded with two aluminum wires on each of the two terminals on the device to four separate low-frequency channels of a QDevil QBoard daughterboard. This allows us to perform four-probe measurements of the differential conductance. 
%If the line resistance is known to be low compared to the tunneling resistance, it may be an advantage to use a 2-probe measurement for simplicity. 

As shown in Fig.~\ref{fig:1}a, the four low-frequency wires pass through the various stages of the dilution refrigerator, including a QDevil QFilter (a setup without filter was also tested; both results are included), and eventually terminate at four separate BNC connectors on the room-temperature breakout box. BNC1 is connected to a voltage divider: the amplitude of one input is divided by a network of resistors by $\sim 10^2$; this input is connected to a QDevil QDAC which controls the DC bias voltage. The amplitude of the other input of the voltage divider is divided by $\sim10^5$; this input is connected to the sine out of an SR800-series Stanford research systems lock-in amplifier (we used the particular models SR830, SR860 and SR865); in this way a small-amplitude AC excitation is added to the DC signal generated by the QDAC. 

BNC2 and BNC3 in Fig.~\ref{fig:1}a are connected to the inputs of two parallel Li-75A differential amplifiers. One amplifier is in the DC-coupled setting, and is connected to a Keysight 34465A digital multimeter, measuring the DC bias voltage $V$ across the CBT device (independent of line resistance). The other Li-75A is in the AC-coupled setting and connected to the second lock-in amplifier, measuring the excitation voltage amplitude  $\delta V$ across the CBT device (also independent of line resistance). 

At last,  BNC4 is connected to an Ithaco 1211 current pre-amplifier whose unfiltered output ('X1') is connected to the input of the lock-in amplifier, which thus measures the excitation current $\delta I$ through the CBT device. The ratio of the signals measured in the two lock-in amplifiers constitutes the measured differential conductance $\delta I/\delta V$. 

An out of plane magnetic field of $\geq$50 mT is used to turn off the superconductivity of the aluminum islands, with higher field strengths not expected to cause any thermometer degradation \cite{Pekola2004}.

\subsection*{Equipment settings}\label{subsec:equipset}
Reliable results have been achieved with a lock-in reference frequency in the range between 24 Hz and 130 Hz, a time constant of 100 ms and filter roll-off of 24 dB/oct. In addition, we found it advantageous to apply synchronous filtering to remove the signal at twice the reference frequency without increasing the time constant. The trade-off of increasing the time constant and filter-roll-off is that the response time increases; for these settings the time it takes for the output to settle within 1\% of the final value in response to changes of the input is $\sim$1.2 s.

Choosing the AC-signal amplitude is a balance between the signal-to-noise ratio and the systematic errors associated with the smoothening of the conductance curve due to a large amplitude. Our results were obtained with $\delta V$ of the order of a few $\mu V$. 

\section{Experimental procedure}\label{sec:expproc}
Data acquisition proceeds by sweeping the bias voltage $V_{DAC}$ using the QDAC, while storing values of $V$ measured by the Keysight 34465A digital multimeter, $\delta V$ measured by the first lock-in amplifier and $\delta I$ 

\begin{tcolorbox}[title=Side note: SIN thermometry]\label{sidenote}
\vspace{2mm}
~A superconductor-insulator-normal metal (SIN) junction can be used as a (primary) electron thermometer. When the absolute value of the bias $eV$ is smaller than the superconducting gap ($\Delta\approx 200$~$\mu$eV for aluminum) the current flow $I$ is vanishingly small. When the bias approaches the gap, the expected thermally activated $IV$ characteristics is $I\sim e^{(eV-\Delta)/k_BT_e}$ \cite{SIN}.
This can be rearranged to give the temperature in terms of the slope of the $IV$ curve in a semilog plot:
\begin{equation}\label{EqSIN}
    k_BT_e=\frac{d(eV)}{d \text{ln}(I)}.
\end{equation}
An example of fitted $T_e$, along its underlying $IV$ data, is shown in Fig.~\ref{fig:2}. Contrary to the CBT device, the SIN device can only operate in small magnetic fields, since a hard superconducting gap is required~\cite{SIN}.

\vspace{5mm}
\centering
\includegraphics[width=1.00\linewidth]{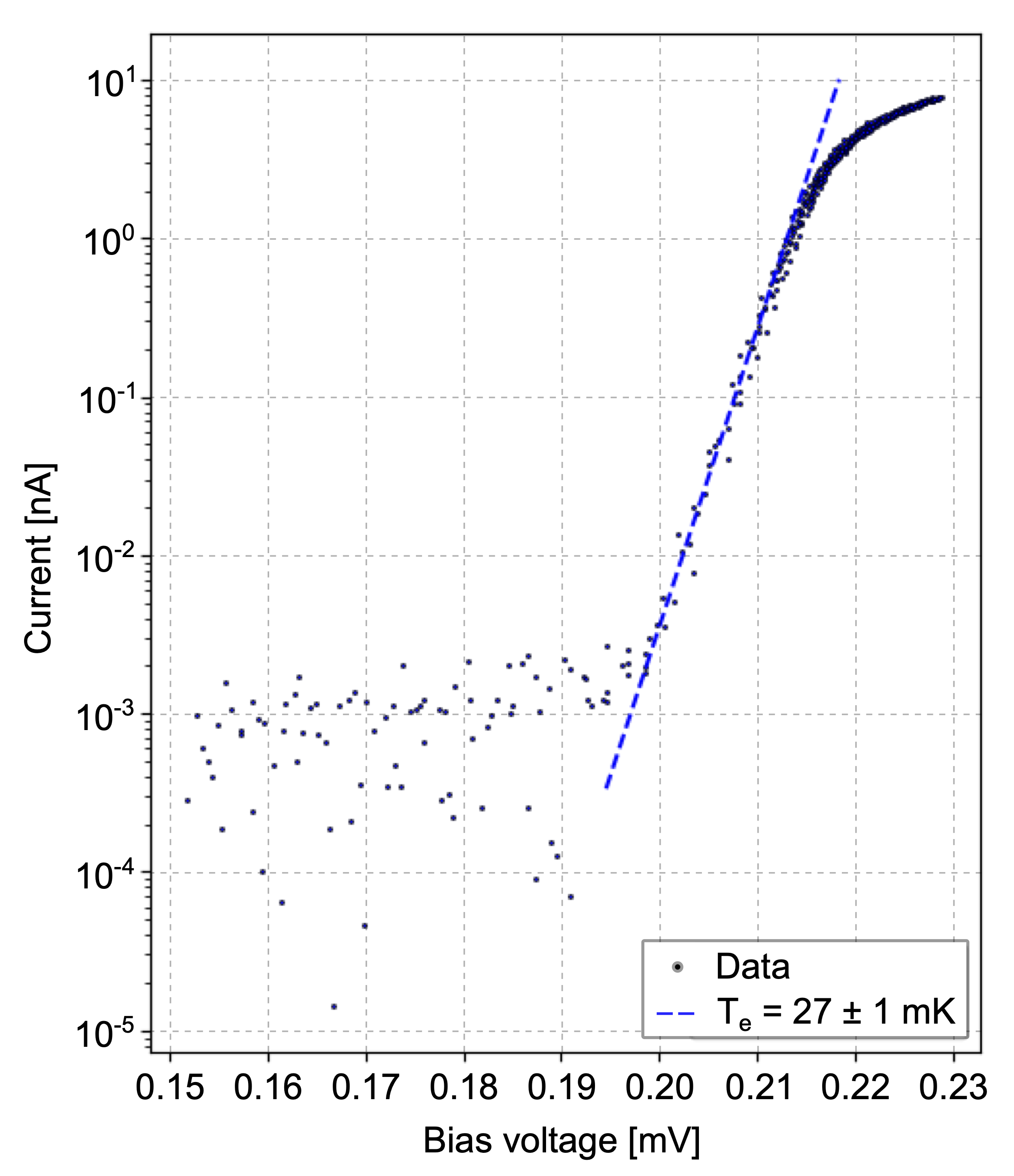}
\captionof{figure}{\textit{Current through a SIN junction as a function of bias voltage around the super-conducting gap $\sim$ 200$~\mu$V, performed in a Bluefors XLD cryostat at a base temperature of 8.8 mK and with the use of QDevil's QFilter. The inverse slope in the illustrated interval is used to extract an electron temperature of $T_e =27\pm 1$ mK. The uncertainty is obtained by repetition measurements.}}\label{fig:2}
\vspace{2mm}
\end{tcolorbox}

\begin{figure*}[!t]
    \centering
    \includegraphics[width=1.00\linewidth]{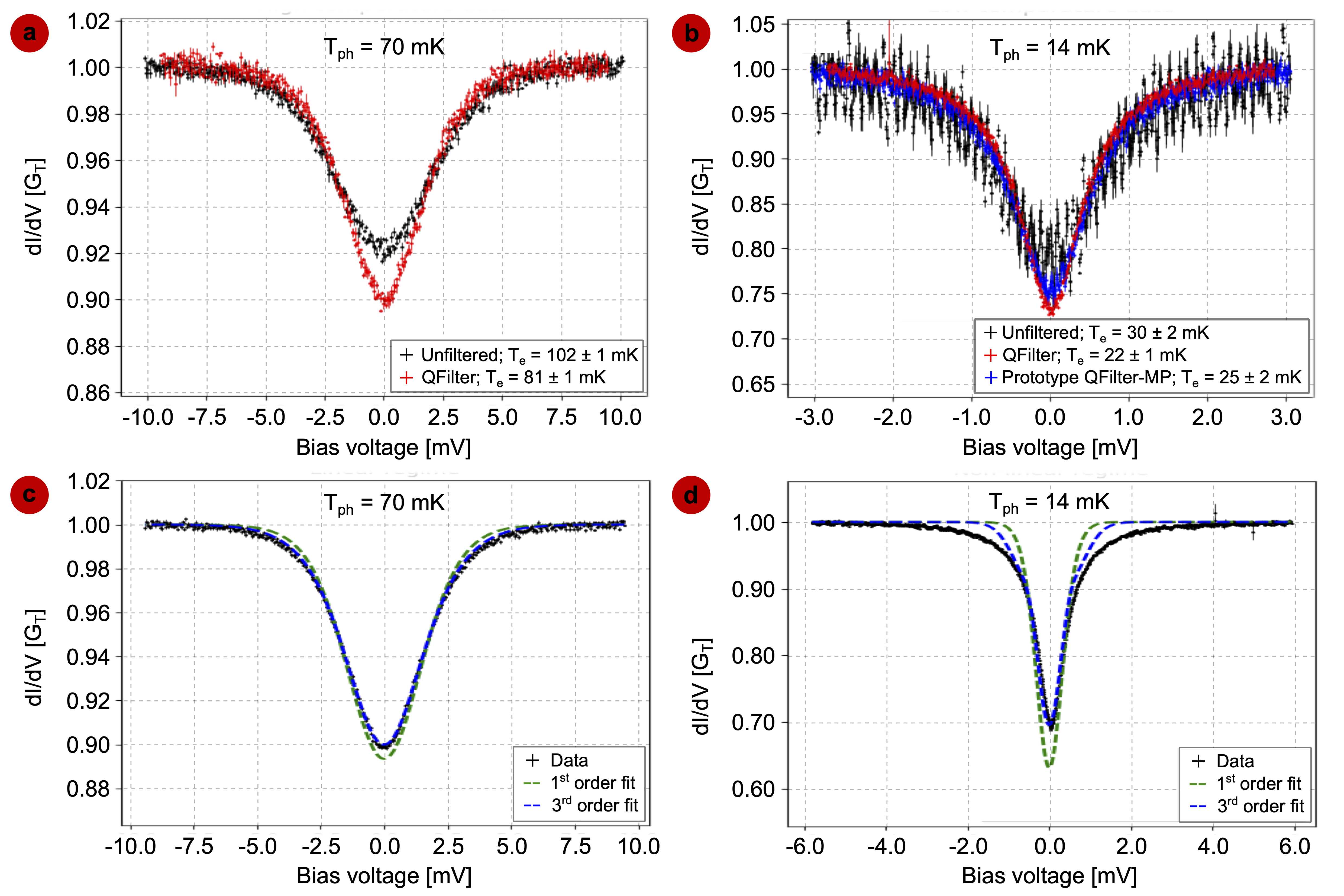}
    \caption{\textit{(a) Two sets of CBT differential conductance data from a dilution refrigerator with a phonon temperature of $\sim$70 mK near the samples; one used a QDevil QFilter (red) and one without the use of any filter (black). (b) Three sets of CBT differential conductance data from a dilution fridge with a phonon temperature of $\sim$14 mK near the samples; one used a QDevil QFilter (red), one without the use of any filter (black), and one with the use of a prototype metal powder QFilter (blue). The temperatures marked in the legend were adjusted for Joule heating (see Appendix \ref{subsec:biasheating}). The uncertainty is found by repetition measurements.
    (c) Theoretical fitting of the QFilter data in (a) with both a $1^{st}$-order (green) and $3^{rd}$-order (blue) approximation (see Appendix \ref{subsec:theory}). (d) The same fitting as in (c) is applied to data similar to the QFilter data in (b) in the temperature regime where these approximations start to break down. See Appendices \ref{subsec:theory}, \ref{subsec:biasheating} and \ref{subsec:sectherm} for further details.}}
    \label{fig:3}
\end{figure*}

measured by the second lock-in amplifier at each step. The necessary sweeping range for the voltage is proportional to the number of metal islands and to the temperature. It is important to choose a range starting and ending near the maximum differential conductance $G_T$ in order to cover the full conductance dip and determine both the FWHM and the depth of the dip with high accuracy. To ensure a good voltage resolution we use a few thousand measurement points in this range.

With the lock-in amplifier settings discussed in section \ref{sec:method}\ref{subsec:equipset}, completely uncorrelated data points can be achieved by using a waiting time of more than a second. However, in our measurements we allow for some correlation between the measurement points and use a waiting time of the same order as the lock-in time constant. With these considerations, results are obtained in the form of the  differential conductance $G=\delta I/\delta V$ as a function of bias voltage $V$, which is used to determine the electron temperature.

Figure~\ref{fig:1}c illustrates how to convert the measured differential conductance into electron temperature. It proceeds as follows: first extract the maximum conductance $G_T$ and the minimum conductance $G_{min}$, which is used to calculate the relative depth of the dip $\Delta G/G_T=(G_T-G_{min})/G_T$. Next, the FWHM $V_{1/2}$ is found by calculating the difference between the points of intersection between the measured conductance curve in units of $G_T$ with the horizontal line at $1-\Delta G/2G_T$. To define $G_T$ and $G_{min}$ a few data points around the minimum and maximum can be averaged, in order to remove the effect of noise. $V_{1/2}$ can be obtained by interpolating the data linearly.

After measuring these quantities, the temperature is given by:
\begin{equation}\label{EqCBT2}
    T_e=\frac{V_{1/2}}{5.439k_BN\left(1+0.3921\frac{\Delta G}{G_T}\right)},
\end{equation}
where $N$ is the number of metallic islands~\cite{heat1997}. Note that Ref.~\cite{intermediate} pointed out a typo in the literature (\cite{Pekola2004}), and accordingly we set the pre-factor of $\frac{\Delta G}{G_T}$ to be indeed 0.3921.

In order to evaluate the CBT method against other thermometry techniques, we compare our CBT results with data from an SIN thermometer (more details of this measurement can be found in the side note).

\begin{figure*}[!t]
    \centering
    \includegraphics[width=1.00\linewidth]{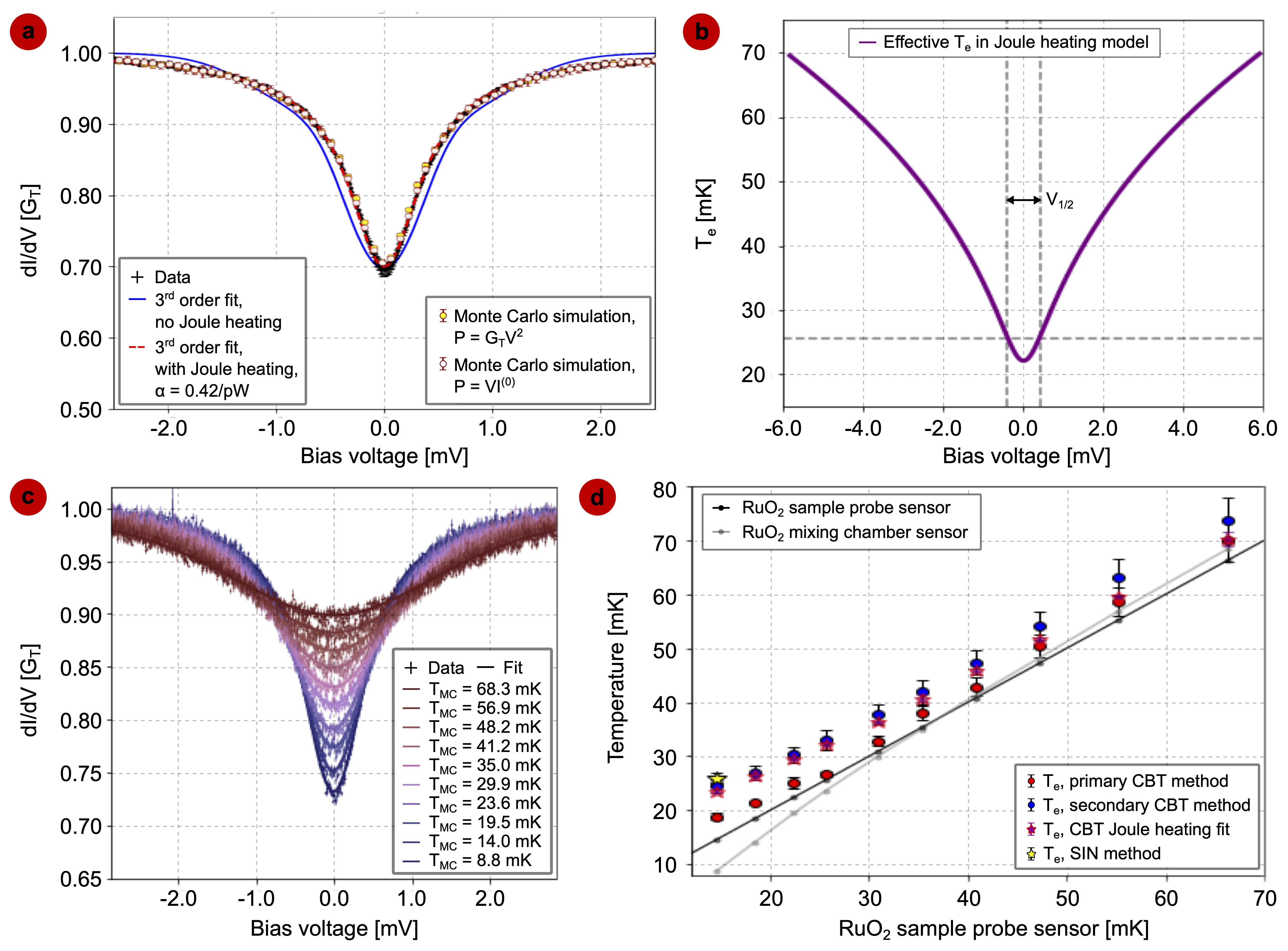}
    \caption{\textit{(a) The same data as in Fig.~\ref{fig:3}d is now compared to the $3^{rd}$-order approximation of the theoretical conductance curve using the electron temperature obtained by the secondary thermometry method (blue line, $T_e = 22\pm 2$mK) (see Appendix \ref{subsec:sectherm}), to a least square fit including the Joule heating effect (dashed red line, $T_0 = 22\pm 1$ mK), with $\alpha \equiv 1/\Sigma\Omega T_0^5$ (see Appendix \ref{subsec:biasheating}), and to numerical Monte Carlo simulations (yellow and white dots) (see Appendix \ref{subsec:MonteCarlo}). (b) The effective electron temperature as a function of bias voltage, given by the Joule heating fit, showing that this effect is significant at $V_{1/2}$, thereby limiting the applicability of equation \ref{EqCBT2}. (c) CBT conductance curves taken at different temperatures of the mixing chamber, plotted together with their theoretical curves corresponding to the Joule heating fits. (d) Comparison of the different CBT temperature reading methods, with the horizontal axis representing the phonon temperature measured resistively in the sample probe. We compare the temperatures as given by the primary reading (equation \ref{EqCBT2}, red points); the temperatures given by the secondary thermometry reading (blue points, see Appendix \ref{subsec:sectherm}); the temperature measured by the SIN thermometer (yellow star); and the temperature fits including the Joule heating effect (purple stars, see Appendix \ref{subsec:biasheating}).}}
    \label{fig:4}
\end{figure*}

\section{Results}\label{sec:results}
Figure~\ref{fig:3}a shows two CBT conductance data sets taken from two devices placed right next to each other in the same daughterboard and cooled down in an Oxford Triton dilution refrigerator with a phonon temperature near the samples measured to be around 70 mK (using a Lake Shore Cryotronics RuO$_2$ sensor). The data set showing the largest dip (red) was obtained using a QDevil QFilter installed in the DC lines, whereas the other data set (black) had no filtering. 

The data sets in Fig.~\ref{fig:3}b show differential conductance obtained in a BlueFors XLD dilution refrigerator with a phonon temperature near the samples measured to be around 14 mK. The largest conductance dip (red) was obtained for the measurement using a QDevil QFilter installed in the DC lines. The blue data set was obtained with a prototype metal powder filter from QDevil, while the black data points were obtained without filtering. 

Note that for all the presented measurements in this application note the low-pass RC filters were removed from the QDevil QBoard sample holder to only compare the effect of the filters under test.

\subsection*{Limitations}
We note that the model in equation \ref{EqCBT2} loses its accuracy for temperatures lower than what corresponds to $\Delta G/G\sim 0.5$ \cite{Pekola2004}, where the device also becomes sensitive to charging effects, therefore limiting its ability to act as a thermometer. The high-temperature limit of operation is determined by how well the dip can be resolved from the background noise, or ultimately by the finite barrier height of the tunnel junctions~\cite{Czech}.

Variations in the microscopic junction properties will lead to deviations from the model presented in equation \ref{EqCBT2}. However, the device is relatively insensitive to such variations; reference \cite{pekola2021influence} found that a 10\% relative rms random variation in tunneling resistances leads to a temperature error of less than 2\%.

In order to verify that the estimated temperatures using equation \ref{EqCBT2} are consistent with the underlying theory, the theoretical conductance curves corresponding to the measured temperatures are plotted in Fig. \ref{fig:3}c and \ref{fig:3}d. The derivations for the $1^{st}$- and $3^{rd}$-order approximations can be found in Appendix \ref{subsec:theory} In Fig.~\ref{fig:3}c we find that in this 'high temperature' regime around 70 mK the $1^{st}$-order approximation gives a good estimate of the data and the $3^{rd}$-order approximation describes the conductance dip very well. In Fig.~\ref{fig:3}d, in the 'low temperature' regime around 20 mK, we find that only the $3^{rd}$-order approximation fits the data well at low bias voltages.

In this latter case the deviations at higher bias voltages are caused by the effect of Joule heating. This effect is caused by insufficient heat transport away from the metallic islands, which depends on their size, and leads to a voltage dependent electron temperature. The Joule heating sets a limit of applicability for equation \ref{EqCBT2} for devices with relatively small island sizes~\cite{heat1997}. Our observation of the Joule heating effect is consistent with independent measurements of the same device design \cite{OurSensor2011}. Newer generations of devices have already proven to be less affected by Joule heating \cite{radioFreq,samani2021microkelvin,JoonasRef1,JoonasRef2}, even in the sub-millikelvin regime~\cite{JoonasRef3}. Further details can be found in Appendix~\ref{subsec:biasheating}

It is however possible to infer the temperature from the CBT data in a way that is independent of the Joule heating, and only depends on the depth of the conductance dip $\Delta G/G_T$. This requires extra knowledge of the device, namely the charging energy of the islands, which can either been known in advance or calibrated from a measurement that is not much affected by the Joule heating effect. (This additional calibration makes this a secondary thermometry method.) The details of this method are outlined in Appendix \ref{subsec:sectherm}. 

Figure~\ref{fig:4}a compares the theoretical conductance curves with and without Joule heating, showing that including this effect provides a reasonable explanation for the found deviations. Figure~\ref{fig:4}b shows that the minimum electron temperature (at zero bias voltage) nevertheless agrees with the value obtained by the secondary thermometry method. Additionally, Monte Carlo simulations (similar to \cite{Yur, numericalPekola}) are used in Fig.~\ref{fig:4}a to verify the Joule-heating-corrected fit, and to show that higher order corrections in $E_c/k_BT_e$ and offset-charging effects play an insignificant role for these parameters. For more details, see Appendix \ref{subsec:MonteCarlo}.

To compare all thermometry methods applied in this study, we measured the CBT conductance curves while adjusting the fridge temperature by applying a heat load between 0 and 150 $\mu$W at the mixing chamber plate, as shown in Fig.~\ref{fig:4}c. Figure~\ref{fig:4}d shows the comparison between the primary method (red dots, equation \ref{EqCBT2}), the secondary method (blue dots, equation \ref{EqTheory1} in Appendix \ref{subsec:theory}), the Joule-heating-corrected fitting (purple stars), and the SIN measurement (yellow star). For the secondary reading, the charging energy was extracted at an electron temperature around $80$ mK. We observe that the secondary method generally yields higher temperatures and that it gives a result consistent with the SIN measurement at the lowest temperature.

\section{Conclusion}\label{sec:conclusion}
Coulomb blockade thermometry is a practical technique for measuring the electron temperature in the millikelvin regime. We have described in detail how to set up an experiment with a four-probe CBT measurement using lock-in amplifiers and presented multiple approaches to the temperature analysis and its limitations.

This thermometry technique has allowed us to characterize the performance of QDevil's filters as compared to unfiltered DC lines. The use of a QDevil QFilter leads to both a significant reduction of the electron temperature as well as a better noise performance, and allows for the electrons to be cooled to temperatures below 25 mK.

For our devices we observed a systematic error on the primary temperature reading, using equation \ref{EqCBT2}, at temperatures below 50 mK, which we ascribed to the effect of Joule heating. This observation is based on the comparison with SIN thermometry and theoretical conductance curves, and it is consistent with independent measurements of the same device design.

\section*{Acknowledgments}\label{sec:acknowledgments}
We are grateful to Jukka Pekola and Joonas Peltonen for providing the CBT and SIN samples and insightful discussions about data and analysis. We also thank Merlin von Soosten, S\o ren Andresen and Fabio Ansaloni for their help with performing the measurements and preparing this application note.

\section*{Appendices}\label{sec:Appendices}

\subsection{Theoretical conductance curves}\label{subsec:theory}

To find the theoretical conductance curve of a CBT, next to the number of islands $N$ and the electron temperature, the charging energy of the islands is required. The ratio between the thermal energy and the charging energy, $\frac{1}{u}=\frac{k_BT_e}{E_c}$, depends only on $x\equiv \Delta G/G_T$ \cite{intermediate}, and is given by:
\begin{equation}\label{EqTheory1}
    \frac{k_BT_e}{E_c}=\frac{1}{6x}-\frac{1}{10}-\frac{1}{350}x+\frac{27}{875}x^2.
\end{equation}
The theoretical current $I(V)$ can be calculated from the master equation to the third order in $u$~\cite{Pekola2004}:
\begin{equation}\label{EqTheory2}
    \frac{R_TI(V)}{Nk_BT_e}=v-uf-\frac{u^2}{4}f''h-\frac{u^3}{8}\left[\frac{1}{4}f''''h^2+\frac{1}{3}f''\right],
\end{equation}
where $v=eV/(Nk_BT_e)$, $R_T=1/G_T$ and $f^{(i)}=f^{(i)}(v)$ is the $i$th derivative of:
\begin{equation}\label{EqTheory3}
    f(v)=\frac{1}{2}-\frac{ve^v-e^v+1}{(e^v-1)^2},
\end{equation}
and $h=h(v)$ is given by:
\begin{equation}\label{EqTheory4}
    h(v)=\frac{v}{\text{tanh}(v/2)}.
\end{equation}
The conductance curve is then given by the derivative of equation \ref{EqTheory2}; $G(V)=\frac{dI}{dV}$. In many cases the first order, linear approximation gives a satisfactory expression:
\begin{equation}\label{EqTheory5}
    G(V)/G_T=1-uf'(v)+\mathcal{O}(u^2),
\end{equation}
with:
\begin{equation}\label{EqTheory6}
    f'(v)=\frac{v\text{sinh}(v)-4\text{sinh}^2(v/2)}{8\text{sinh}^4(v/2)}.
\end{equation}
Higher order corrections up until third order can be calculated from equation \ref{EqTheory2}. It is expected that any contributions beyond third order will be negligible compared to other effects~\cite{Pekola2004}.

Note that for numerical implementation, it may be necessary to use several of the first terms in the Taylor expansion of $f(v)\approx v/6-v^3/180+v^5/5040-v^7/151200$ and $h(v)\approx2+v^2/6-v^4/360$ and their derivatives when evaluating the expressions around $v=0$.

\subsection{Joule heating}\label{subsec:biasheating}
The bottleneck for transporting heat away from the metallic islands is in general the small electron-phonon thermal coupling. In steady state, the heat balance can be written as~\cite{Wellstood}:
\begin{equation}\label{EqHeating1}
    T_e^5-T_0^5=\frac{P}{\Sigma\Omega},
\end{equation}
where $T_e=T_e(V)$ is the effective electron temperature when a bias voltage $V$ is applied, $P=I(V)V\approx G_TV^2$ is the dissipated power, $\Omega$ is the total volume of the metallic islands, $\Sigma$ is a material dependent parameter (for aluminum $\Sigma=0.4\pm0.1\times10^9\frac{W}{K^5m^3}$~\cite{Pekola2004}), and $T_0$ is the temperature at zero bias.

The modified conductance curve in the presence of Joule heating can be calculated by differentiating equation \ref{EqTheory2}, while treating $T_e$ as a voltage-dependent parameter:
\begin{equation}\label{EqHeating2}
    T_e(V)=\left[\frac{P}{\Sigma\Omega}+T_0^5\right]^{1/5}.
\end{equation}
In Fig.~\ref{fig:4}a a least squares fit to experimental data with respect to $\Sigma\Omega$ and $T_0$ is shown.

\subsection{Secondary CBT thermometry}\label{subsec:sectherm}

As the ratio $\Delta G/G_T$ of a CBT conductance dip is unaffected by the Joule heating effect, it can be utilized for electron thermometry. To do so, however, knowledge of the islands charging energy is required. Combining equation \ref{EqTheory1} with equation \ref{EqCBT2} allows us to extract both $E_c$ and $T_e$ from $\Delta G/G_T$ and $V_{1/2}$. Note that $E_c$ is specific to the device. Once $E_c$ is measured at a temperature where Joule heating plays a negligible role, the lower electron temperatures can be extracted from using solely equation \ref{EqTheory1} which does not depend on $V_{1/2}$ and therefore has the advantage that it is not affected by the Joule heating effect discussed in Appendix \ref{subsec:biasheating}

\subsection{Monte Carlo simulations}\label{subsec:MonteCarlo}
In Fig.~\ref{fig:4}a we compare the Joule-heating-corrected fit with Monte Carlo simulations, following previous work~\cite{Yur, numericalPekola}. As input for the simulations we use the zero-bias temperature obtained from the secondary thermometry method and the Joule heating parameter obtained from the $3^{rd}$-order least square fit with Joule heating. In these simulations a random offset-charge distribution in the interval $\pm e$ is assumed. For the first simulation (yellow dots in Fig.~\ref{fig:4}a) the effective temperature is assumed to be $(T_0^5+G_TV^2/\Sigma\Omega)^{1/5}$, indicating consistency with the third-order fitting result. Another Monte Carlo simulation (white dots in Fig.~\ref{fig:4}a) uses an effective temperature given by $(T_0^5+VI^{(0)}(V)/\Sigma\Omega)^{1/5}$, where $I^{(0)}(V)$ is the current obtained from the former simulation. This suggests that $P\approx G_TV^2$ is a reasonable approximation, since the two Monte Carlo simulations give similar results.

\subsection{Software}\label{subsec:software}
The QCoDeS data acquisition framework provides useful software tools for performing the measurement in Python (see \url{qcodes.github.io}) \cite{qcodes}.
Python scripts for data collection and data analysis may be provided upon request.

\bibliography{bibliography/mybibfile}

\end{document}